\documentclass[a4paper,11pt]{elsarticle}
\usepackage{graphicx, amssymb, amsfonts,amsmath}
\usepackage{multirow}
\newcommand{\rmo}{_{\rm O}}
\newcommand{\rme}{_{\rm E}}

\long\def\symbolfootnote[#1]#2{\begingroup%
\def\thefootnote{\fnsymbol{footnote}}\footnote[#1]{#2}\endgroup}

\begin{document}

\begin{frontmatter}
\title{{\footnotesize\vspace*{-2cm}\hspace*{10cm}DESY 12-223}\\\vspace*{1cm}Lattice Hamiltonian
approach to the massless Schwinger model: precise extraction of the
mass gap}

\author[label1,label2]{Krzysztof Cichy\corref{cor}}
\ead{krzysztof.cichy@desy.de}
\author[label2]{Agnieszka Kujawa-Cichy}
\ead{kujawa@amu.edu.pl}
\author[label2,label3]{Marcin Szyniszewski}
\ead{mszynisz@gmail.com}

\cortext[cor]{Corresponding author. Tel.: +49 33762 77306; fax: +49 33762 77419.}
\address[label1]{NIC, DESY, Platanenallee 6, D-15738 Zeuthen, Germany}
\address[label2]{Adam Mickiewicz University, Faculty of Physics,\\
Umultowska 85, 61-614 Poznan, Poland}
\address[label3]{NOWNano DTC, University of Manchester, Manchester, M13 9PL, UK}

\begin{abstract}
We present results of applying the Hamiltonian approach to the massless Schwinger model. A
finite basis is constructed using the strong coupling expansion to a very high order. Using exact
diagonalization, the continuum limit can be reliably approached. This allows to reproduce the
analytical results for the
ground state energy, as well as the vector and scalar mass gaps to an outstanding precision better
than $10^{-6}$\,\%.
\end{abstract}



\end{frontmatter}

\bibliographystyle{elsarticle-num}

\section{Introduction}
The Schwinger model \cite{Schwinger:1962tp}, i.e. quantum electrodynamics in 1+1 dimensions, is the
simplest gauge theory. 
Since its formulation in 1962, it has attracted much attention. Notwithstanding its apparent
simplicity, its physics is surprisingly rich and in several aspects resembles much more complex
theories, in particular quantum chromodynamics (QCD).
As such, the Schwinger model has become the standard toy model for testing lattice
techniques\footnote{See e.g.
Refs.~\cite{Gutsfeld:1999pu,Gattringer:1999gt,Giusti:2001cn,Christian:2005yp,Bietenholz:2011ey} and
references cited therein.}.

In particular, it was proposed to use lattice Hamiltonian methods for investigation of its
properties
\cite{Banks:1975gq,Susskind:1975hj,Carroll:1975gb,Kenway:1977dk,Crewther:1979ka,Jones:1979av,
Hamer:1982mx,Irving:1982yw,Schiller:1983sj,Fang:1992bi,Aroca:1997hp,Hamer:1997dx,Kroger:1998se,
Sriganesh:1999ws,Fang:2001gq,Byrnes:2002nv}.
Using these techniques, several properties of the massless and massive Schwinger model were
investigated and many interesting results were obtained.

In this paper, we concentrate on the massless case for the 1-flavour model.
Our main aim is to show that lattice Hamiltonian methods can yield results with unprecedented
precision -- \emph{a few orders of magnitude more precise} than their previous applications.
We concentrate on three simple quantites -- the ground state (GS) energy, the scalar
mass gap and the vector mass gap.

In Sec.~\ref{sec:model} we introduce the essentials of the Hamiltonian approach and the lattice
method used.
Sec.~\ref{sec:results} presents our results.
In Sec.~\ref{sec:summary}, we summarize, compare our results with previous estimates and shortly
discuss the prospects of the method.

\section{Model setup}
\label{sec:model}
We start with the Hamiltonian of the lattice Schwinger model in the Kogut-Susskind (staggered)
discretization \cite{Kogut:1976,Banks:1975gq}:
\begin{align}
\label{staggered}
\mathcal{H}=-\frac{i}{2a}\sum_{n=1}^M
\Big(\phi^\dagger(n) e^{i\theta(n)}\phi(n+1) - \phi^\dagger(n+1)e^{-i\theta(n)}
\phi(n)\Big)+\nonumber\\
+\,m\sum_{n=1}^M (-1)^n\phi^\dagger(n)\phi(n)+\frac{ag^2}{2}\sum_{n=1}^M
L^2(n),
\end{align}
where $\phi(n)$ is a single-component fermion field\footnote{The staggered discretization can be
viewed as starting with a two-component fermion field on each lattice site and ignoring the upper
component on odd sites and the lower component on even sites. In this way, one avoids the fermion
doubling
problem and obtains a well defined continuum limit in physical observables.}, defined on
each site of an $M$-site lattice with periodic boundary conditions (i.e. on a circle) and obeying the
anticommutation relations $\{\phi^\dagger(n),\phi(m)\}=\delta_{nm}$, $\{\phi(n),\phi(m)\}=0$,
$\{\phi^\dagger(n),\phi^\dagger(m)\}=0$. $m$ denotes the fermion (``quark'') mass. The gauge field
variable $\theta(n)$ is defined on the link between sites $n$ and $n+1$ and is related to the spatial
component of the Abelian vector potential by $\theta(n)=agA(n)$ (we work in the temporal gauge
$A_0=0$), where $g$ is the gauge coupling and $a$ is the lattice
spacing. The variable $L(n)$ is related to the electric field $E(n)$ by the
relation $L(n)=E(n)/g$ and to the gauge field by the commutation relations:
$[\theta(n),L(m)]=i\delta_{nm}$. The possible values of $L(n)$ are quantized:
$L(n)|l\rangle=l|l\rangle$, where $l=0,\pm1,\pm2,\ldots$.
This implies the following action of $e^{\pm i\theta(n)}$ on the basis states: $e^{\pm
i\theta(n)}|l\rangle=|l\pm1\rangle$.

For numerical treatment, it is convenient to perform the Jordan-Wigner transformation \cite{jw}
$\phi(n)=\prod_{p<n}(i\sigma^3(p))\sigma^-(n)$, where $\sigma^i(n)$ are Pauli matrices
($\sigma^\pm=\sigma^1\pm i\sigma^2$). This gives:
\begin{align}
\label{staggered2}
\mathcal{H}_{\rm JW}=\frac{1}{2a}\sum_{n=1}^M
\Big(\sigma^+(n) e^{i\theta(n)}\sigma^-(n+1) + \sigma^+(n+1)e^{-i\theta(n)}
\sigma^-(n)\Big)+\nonumber\\
+\,\frac{m}{2}\sum_{n=1}^M \left(1+(-1)^n\sigma^3(n)\right)+\frac{ag^2}{2}\sum_{n=1}^M
L^2(n).
\end{align}

Let us now consider the choice of the basis, which is essential for numerical investigations.
The natural choice is the direct product of Ising basis, acted upon by Pauli spin operators, and
the ladder space of states \{$|l\rangle$\}, acted upon by the operator $L(n)$ and the rising and
lowering operators $e^{\pm i\theta(n)}$. With an $M$-site lattice, the dimension of the spin part
is $2^M$, while for the gauge part the basis is infinite-dimensional. Hence, the whole basis is
infinite-dimensional even on a finite lattice. Clearly, for numerical computation, some choice of (a
finite number of) basis states has to be made. One possibility, motivated by the physics of the
problem, is to
truncate at some finite $\pm l_{\rm max}$. However, such an approach gives prohibitively large bases
even
for moderate lattice sizes $M$, since the region of physically important values of $l$ quickly
increases as one approaches the
continuum limit, adding to the exponential increase of the basis size from the spin part.

Fortunately, a much better truncation procedure exists, using the strong coupling expansion (SCE)
\cite{Kogut:1976,Banks:1975gq}. Rewrite the Hamiltonian in a dimensionless form:
\begin{equation}
 W=\frac{2}{ag^2}\mathcal{H}_{\rm JW}=W_0+xV,
\end{equation} 
with:
\begin{equation}
W_0=\frac{m}{ag^2}\sum_{n=1}^M \left(1+(-1)^n\sigma^3(n)\right)+\sum_{n=1}^M L^2(n),
\end{equation} 
\begin{equation}
V=\sum_{n=1}^M \Big(\sigma^+(n) e^{i\theta(n)}\sigma^-(n+1) + \sigma^+(n+1)e^{-i\theta(n)}
\sigma^-(n)\Big)
\end{equation} 
and
\begin{equation}
x\equiv\beta=1/a^2g^2. 
\end{equation}  
The continuum limit of the model corresponds to $x\rightarrow\infty$, $a\rightarrow0$,
$M\rightarrow\infty$. The SCE parameter $x$ is conventionally denoted $\beta$ in lattice gauge
theory literature.

Formally, the operator $W_0$ can be treated as an unperturbed part and $V$ as a perturbation. It is
easy to see that the ground state of $W_0$, which we will denote by $|0\rangle$, is given by:
\begin{equation}
 |0\rangle =
|\!\downarrow\uparrow\downarrow\uparrow\cdots\downarrow\uparrow\rangle\otimes|0000\cdots00\rangle,
\end{equation} 
i.e. ``antiferromagnetic'' spin state and no gauge field excitations ($L(n)=0$ for all sites $n$),
$W_0|0\rangle=0$.
The perturbation operator $V$ flips two neighbouring spins\footnote{Hence, all generated basis states
have zero total magnetization $\sum_n \sigma^3(n)$ for the spin part.} and couples them via
a gauge field
excitation (flux line) $L(n)=\pm1$. For example, for a 4-site lattice:
\begin{equation}
 V|0\rangle=|\!\downarrow\uparrow\uparrow\downarrow\rangle\otimes|000{\textrm -}\!1\rangle+
|\!\downarrow\downarrow\uparrow\uparrow\rangle\otimes|0010\rangle+
|\!\uparrow\downarrow\downarrow\uparrow\rangle\otimes|0{\textrm -}\!100\rangle+
|\!\uparrow\uparrow\downarrow\downarrow\rangle\otimes|1000\rangle.
\end{equation} 
Conventionally, a fermionic excitation on an odd site (subscript O; the sites are numbered from
right to left, $n=1,2,\ldots,M$), i.e. $|\!\downarrow\rmo\rangle$
($|\!\uparrow\rmo\rangle$ means no excitation, i.e. spin alignment corresponding to the ground
state) is referred to as a ``quark'', whereas an excitation on an even site (subscript E), i.e.
$|\!\uparrow\rme\rangle$ is an ``antiquark''. Thus,
$|\!\uparrow\rme\downarrow\rmo\rangle\otimes|0{\textrm -}\!1\rangle$ is a ``meson''\footnote{We
use the standard convention of naming the fermions in the Schwinger model ``quarks'' or
``antiquarks'' and their pairs (connected by flux lines) ``mesons'' or ``antimesons''. Of course,
these particles have nothing to do with real-world quarks and mesons of QCD.} (quark-antiquark
with a directed gauge field link connecting them) and, correspondingly
$|\!\downarrow\rmo\uparrow\rme\rangle\otimes|10\rangle$ an ``antimeson'' (flux line directed in the
other direction).

The above state $V|0\rangle\equiv|1^S\rangle$ is hence a superposition of all possible 1-meson
(``scalar'' -- superscript $S$) states corresponding to one quark -- one antiquark excitation. We
will call such a superposition a \emph{class of states}. A class of states is translationally
invariant, however, since we are working in the staggered lattice formalism, a meson can only be
translated by an even number of sites, if an odd number of sites translation is performed, the meson
is transformed to an antimeson (``charge conjugation'' operation). One additional symmetry operation
that can be performed on each class of states is a change of ``helicity'' $H$ i.e. the clockwise
numbering of lattice sites to an anti-clockwise (and taking into account the inequivalence of even
and odd sites), or vice versa.
Any of the following operations, or any combination of them -- translation by two lattice sites
$T_2$, ``charge conjugation'' $C$ and ``helicity'' $H$ -- leaves the class of states $|1^S\rangle$
(and any other class) invariant.

The construction of the dimensionless Hamiltonian $W$ basis is performed in the following way.
One starts with the ground state $|0\rangle$ and acts with the perturbation operator $V$ on it to
obtain a 1-meson class of states $|1^S\rangle$. Then, one acts again with $V$, obtaining a 2-meson
class of states and a 0-meson class of states (i.e. ground state) etc. In general, a $k$-meson
class of states, acted on with $V$, can produce several $(k+1)$- and $(k-1)$-meson classes of
states:
\begin{equation}
\label{eq:states}
 V|k_i^S\rangle = \sum_j a_j |(k+1)_j^S\rangle + \sum_j b_j |(k-1)_j^S\rangle,
\end{equation} 
where $|k_i^S\rangle$ is $i$-th $k$-meson class of ``scalar'' states and $a_j$, $b_j$ are
coefficients that have to be calculated for each state. States belonging to a given class of states
are all states related by the operations $T_2$, $C$ and $H$.
All generated classes of states are eigenstates of the unperturbed operator $W_0$ (diagonal
elements of $W$), while states with different numbers of mesons are related via off-diagonal matrix
elements $\langle k_i^S|V|(k+1)_j^S\rangle$.

One can also construct the Hamiltonian $W_V$ starting from the lowest ``vector''
(1-meson) class of states (instead of the 0-meson state $|0\rangle$):
\begin{equation}
|1^V\rangle =  \frac{1}{\sqrt{M}}\sum_n \left(\sigma^+(n) e^{i\theta(n)}\sigma^-(n+1) -
\sigma^+(n+1)e^{-i\theta(n)}\sigma^-(n)\right)|0\rangle.
\end{equation} 
Acting several times with $V$ on $|1^V\rangle$ allows to obtain the ``vector''-states Hamiltonian
matrix $W_V$ in the $|k_i^V\rangle$ basis.

Obviously, the construction of the Hamiltonian basis can proceed indefinitely, since each site can
be connected to a neighbouring one by gauge links (flux lines) with arbitrary $|L(n)|=0,1,2,\ldots$.
Therefore, we truncate at some finite order $N$ of SCE. We will show that a finite $N$ is enough to
extract the relevant physics. The value of $N$ needed depends on the lattice spacing and grows as
the continuum limit is approached.
A particular role, to be discussed below, is played by closed flux lines, i.e. flux loops. When the
order of SCE $N$ becomes equal to the number of sites $M$, the following class of states appears:
\begin{equation}
\label{eq:loops}
|0_{1-{\rm loop}}\rangle=
|\!\downarrow\uparrow\downarrow\uparrow\cdots\downarrow\uparrow\rangle\otimes|1111\cdots11\rangle +
|\!\downarrow\uparrow\downarrow\uparrow\cdots\downarrow\uparrow\rangle\otimes|{\textrm
-}\!1{\textrm -}\!1{\textrm -}\!1{\textrm -}\!1\cdots{\textrm -}\!1{\textrm -}\!1\rangle,
\end{equation} 
with all spins pointing in the same direction as in the ground state (no quarks or antiquarks and
hence no mesons), but with a flux line connecting all sites and hence forming a flux loop (the sign
of `1' in the ladder space part of a basis state corresponds to either a clockwise or anticlockwise
orientation of gauge links circulating around the loop). Increasing $N$, one encounters states with
an increasing number of flux loops -- Eq.~\eqref{eq:loops} with `1' replaced by an arbitrary $N_{\rm
loop}$ (starting at SCE order $N=N_{\rm loop}M$).

To summarize this part, let us shortly comment on the total size of the basis and its scaling with
the lattice size. In the adopted approach, the size of the spin part of the basis grows much slower
than exponentially.
Zero magnetization sector for an $M$-site lattice consists of $M\choose 2$ states. Grouping states
into classes of states by using symmetry operations $T_2$, $C$ and $H$ further decreases the spin
part basis size.
The total basis size is then determined by the value of the maximal allowed
number of flux loops $N_{\rm loop}$, i.e. by the ratio $N/M\,$.
Note that some given maximal allowed $N_{\rm loop}$ does not imply in general that $l_{\rm max}=N_{\rm
loop}+1$, as can be naively expected\footnote{This is most easily understood by looking e.g. at the
following state: 
$|\uparrow\downarrow\uparrow\uparrow\uparrow\downarrow\downarrow\downarrow\rangle\otimes|0{\textrm
-}\!10{\textrm -}\!1{\textrm -}\!1{\textrm -}\!2{\textrm -}\!1{\textrm -}\!1\rangle$, which does not
have closed flux loops (i.e. it occurs even if max. $N_{\rm loop}=0$), but still has $l=\pm2$ in
the gauge part.}.
Therefore, it is still unavoidable that increasing the lattice size $M$ and keeping fixed the ratio
$N/M$, the total basis size grows (approximately) exponentially.
We will give some explicit values of the total basis size in the next section.

In this paper, we are interested in three quantites: the ground state energy, the scalar
mass gap and the vector mass gap. 
The $i$-th eigenvalue ($i=0,\,1,\,\ldots$) of the Hamiltonian $W$ will be denoted by $\omega_i$. We
define the following quantites (and also include the exact values in the continuum limit,
for the
massless case $m = 0$)\,:
\begin{itemize}
\item GS energy:
\begin{equation}
 E_0=\frac{\omega_0}{2Mx}\qquad \xrightarrow[M\rightarrow\infty]{a\rightarrow0}
\qquad -\frac{1}{\pi} \approx -0.3183098862,
\end{equation} 
\item scalar mass gap:
\begin{equation}
 \frac{M_S}{g}=\frac{\omega_1-\omega_0}{2\sqrt{x}} \qquad
\xrightarrow[M\rightarrow\infty]{a\rightarrow0}
\qquad \frac{2}{\sqrt{\pi}} \approx 1.1283791668,
\end{equation} 
\item vector mass gap:
\begin{equation}
 \frac{M_V}{g}=\frac{\omega_0^V-\omega_0}{2\sqrt{x}} \qquad
\xrightarrow[M\rightarrow\infty]{a\rightarrow0}
\qquad \frac{1}{\sqrt{\pi}} \approx 0.5641895836,
\end{equation} 
where $\omega_0^V$ stands for the lowest eigenvalue of the ``vector'' Hamiltonian $W_V$.
\end{itemize}

\begin{table}[t!]
  \centering
  \begin{tabular}{cccc}
    \hline\hline
    $x\equiv\beta$ &  lattice sizes $M$ & max. $N$ & dimension of $W_V$\\
  \hline
  2500 &  8,\,10,\,12,\,14 & 150 & 3347\\
10000 &  8,\,10,\,12,\,14 & 200 & 4454\\
40000 &  8,\,10,\,12,\,14 & 250 & 5561\\
250000 &  8,\,10,\,12,\,14 & 350 & 7776 \\
1000000 &  8,\,10,\,12,\,14 & 600 & 13304\\
4000000 &  8,\,10,\,12,\,14 & 1200 & 26597\\
    \hline\hline\\
  \end{tabular}
  \caption{Simulation parameters -- inverse coupling $x$, lattice
sizes $M$, maximum order of SCE $N$. We also give the dimension of the ``vector'' Hamiltonian
matrix $W_V$ for our largest lattice size $M=14$.}
  \label{tab:param}
\end{table}

\section{Results}
\label{sec:results}
The parameters of lattices used for this work are presented in Tab.~\ref{tab:param}. We work very
close to the continuum limit and move towards it by changing the inverse coupling $x\equiv\beta$ from
2500
to 4000000\,\footnote{Note that typical values of $\beta$ in Monte Carlo simulations for the
Schwinger model are $\mathcal{O}$(5--10).}. The lattices have $M=8$ to 14 sites. The order of SCE
(denoted by
$N$) is chosen such that its further increase does not change the
results up to machine precision. We also illustrate how the size of the Hamiltonian matrix $W_V$ grows
with increasing $N$, at fixed $M$, giving examples for our largest lattice
size $M=14$.
However, as we have mentioned above, the size of the Hamiltonian matrix grows roughly exponentially
when one increases $M$ at a fixed ratio $N/M$ -- e.g. for $N/M=1200/14$, the dimension of $W_V$ is:
1546, 3603, 9615, 26597 for $M=8,\,10,\,12,\,14$, respectively.

\begin{figure}[t!]
\begin{center}
\includegraphics
[width=0.6\textwidth,angle=270]
{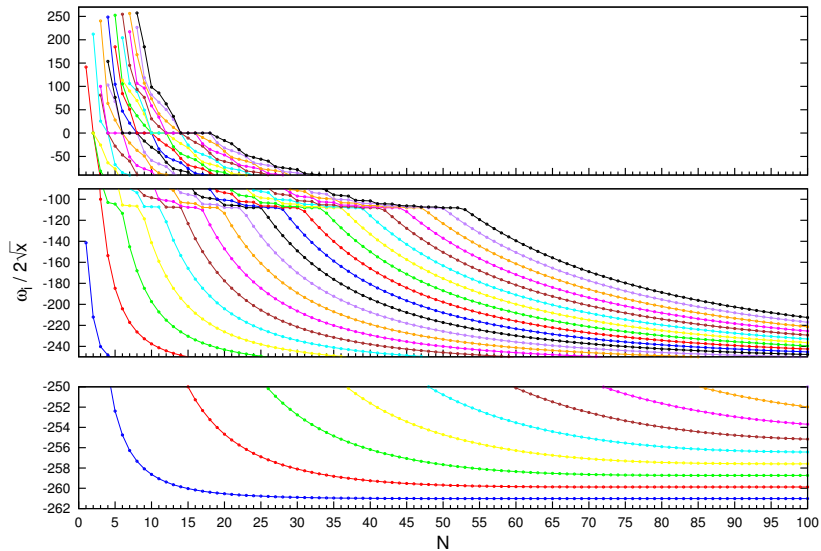}
\end{center}
\caption{\label{fig:EV_flow} Eigenvalue flow with the order of strong coupling expansion $N$. The
vertical scale is split into three regions, for better presentation. $M=8$, $x\equiv\beta=2500$. The
difference between the first excited state energy and the ground state energy is the scalar mass
gap.}
\end{figure}

\begin{figure}[t!]
\begin{center}
\includegraphics
[width=0.34\textwidth,angle=270]
{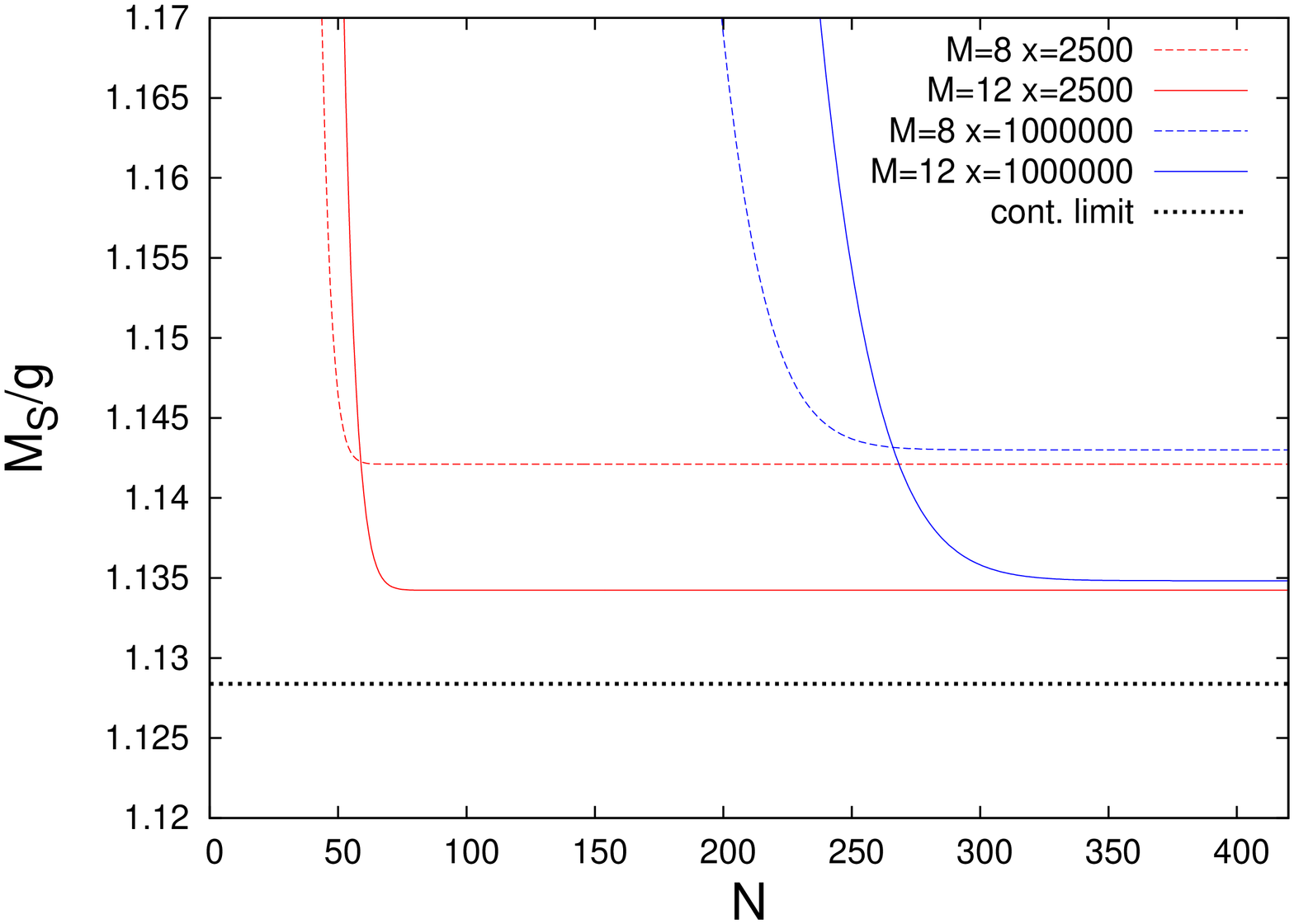}
\includegraphics
[width=0.34\textwidth,angle=270]
{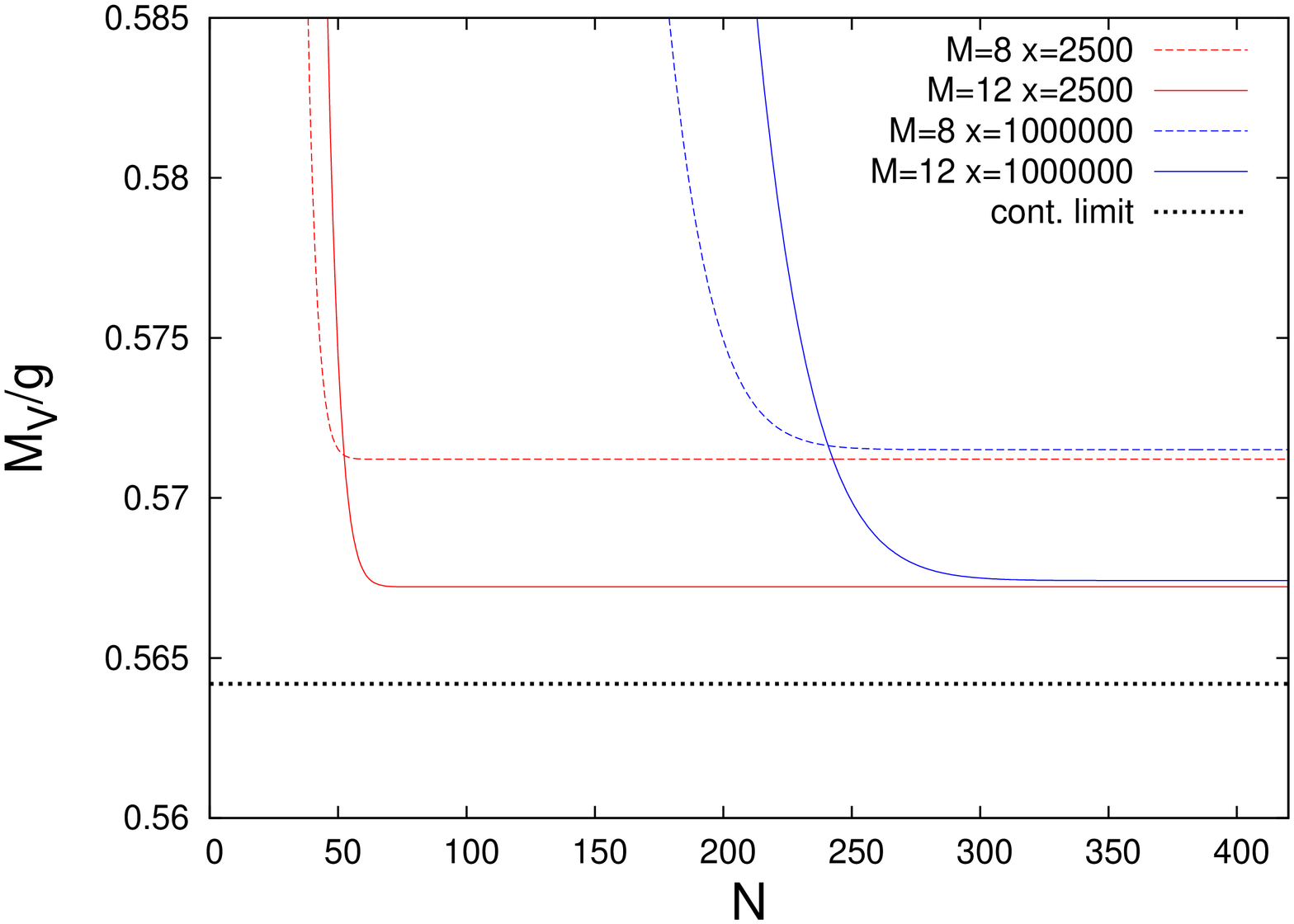}
\includegraphics
[width=0.34\textwidth,angle=270]
{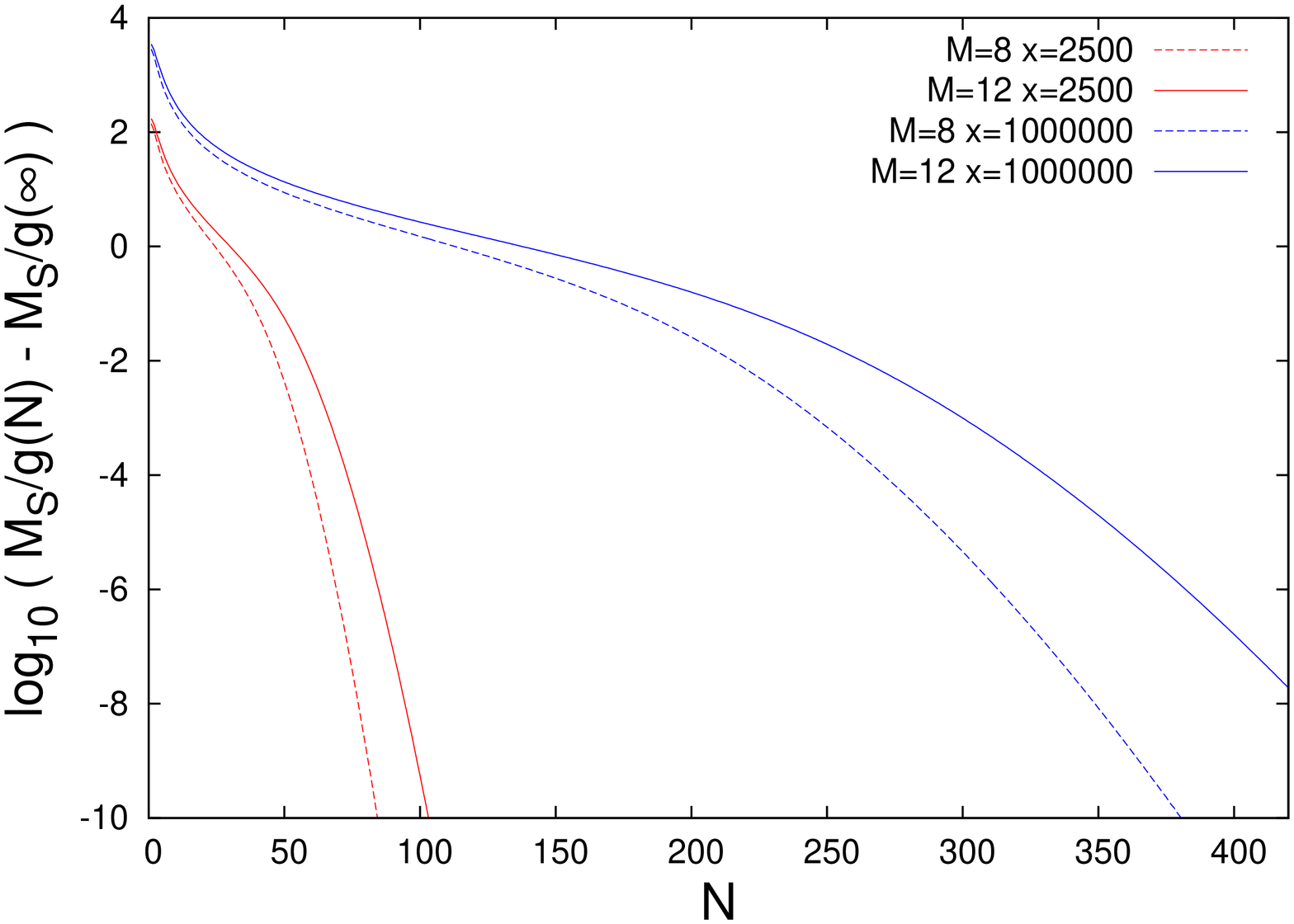}
\includegraphics
[width=0.34\textwidth,angle=270]
{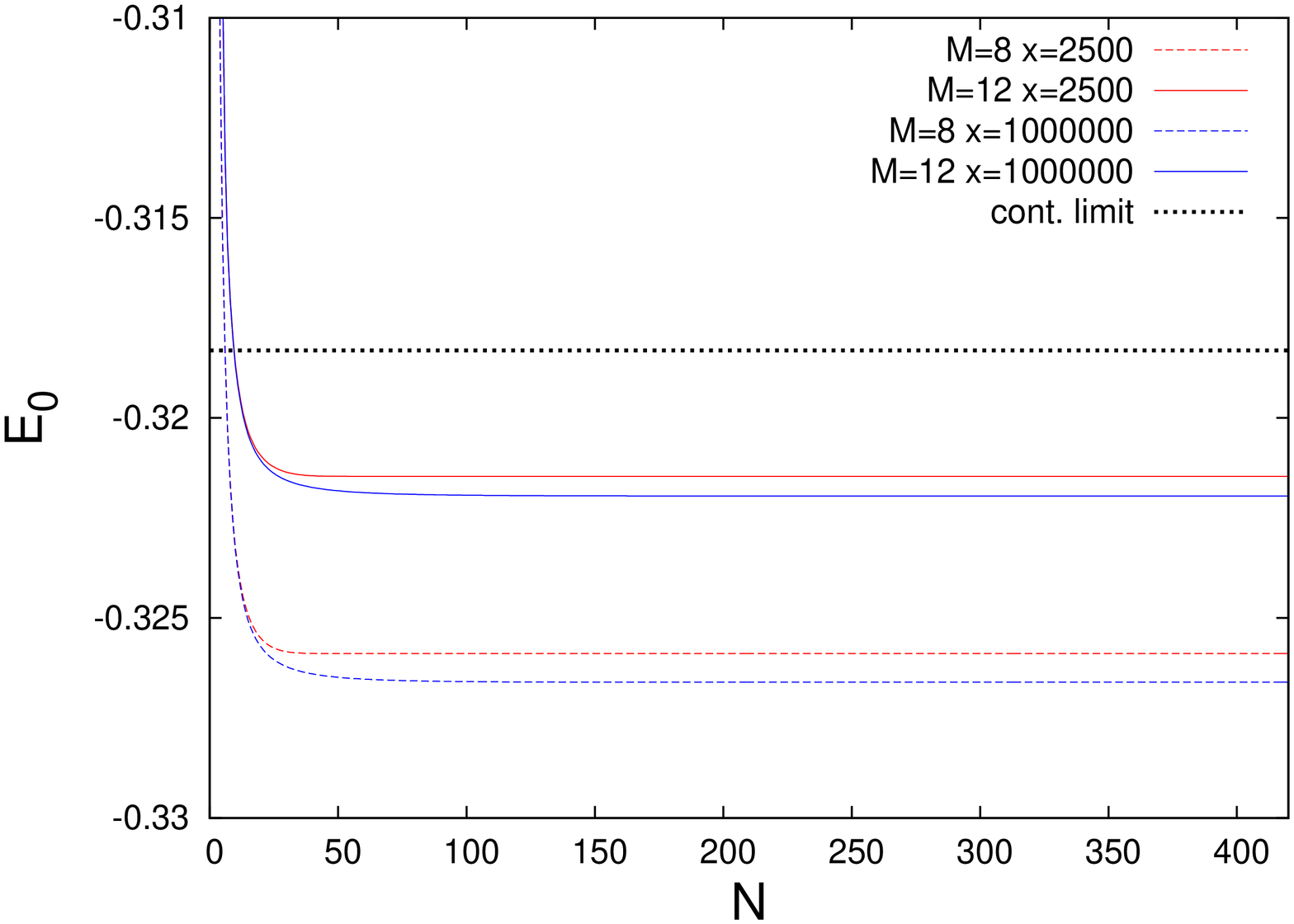}
\end{center}
\caption{\label{fig:flow} Dependence of scalar mass gap (upper left), vector mass gap
(upper right) and ground state energy (lower right). We also show (lower left) the convergence of
the scalar mass gap to the asymptotic ($N\rightarrow\infty$) value. The dotted lines correspond to
$M=8$ and solid lines to $M=12$. Two values of the gauge coupling $x\equiv\beta=2500$ (red) and
$x\equiv\beta=1000000$ (blue) are used. The exact continuum results (black dotted lines) are also
shown.}
\end{figure}

We first discuss the dependence of the results on the order of SCE $N$ and the maximal
allowed number of flux loops taken into account in the computation $N/M$.
In Fig.~\ref{fig:EV_flow}, we illustrate the flow of eigenvalues with $N$ for a small lattice with
8 sites and a rather large coupling $1/\sqrt{x}=0.02$. For better visibility, we split the vertical
scale into three regions.
The most distinctive feature of the plot is that the magnitude of $i$-th eigenvalue is saturated for
some finite value of $N$ which we denote by $N_i$. We observe that $N_i$ increases as $i$ is
increased. Hence, the GS energy is quickly saturated at order $N_1$ (only a few flux loops are
needed), while the scalar mass gap is saturated at a larger value $N_2$. Quantities
involving
higher eigenvalues will, of course, require even higher values of $N$.


This is further illustrated in Fig.~\ref{fig:flow}, where we show how the three quantities of
interest are saturated with increasing $N$. We compare different lattice sizes and different values
of the gauge coupling. It is clear from these plots that the GS energy saturates very rapidly,
while the convergence is not so fast for the mass gaps (it is slightly faster for the vector mass
gap). The observed convergence to the $N\rightarrow\infty$ value is approximately exponential
-- see the lower left plot of Fig.~\ref{fig:flow} for the scalar mass gap case. This plot shows
that the saturation order $N$ for some requested precision is highly dependent on $x$ and $M$. The
contribution of states with an increasing number of allowed flux loops becomes more important as one
moves towards the continuum limit (increasing $x$).
We also observe that at fixed coupling $x$, the saturation order $N$ grows approximately linearly
with $M$, motivating keeping the ratio $N/M$, i.e. the maximal allowed number of flux loops, fixed
when increasing $M$ at fixed $x$.

\begin{figure}[t!]
\begin{center}
\includegraphics
[width=0.6\textwidth,angle=270]
{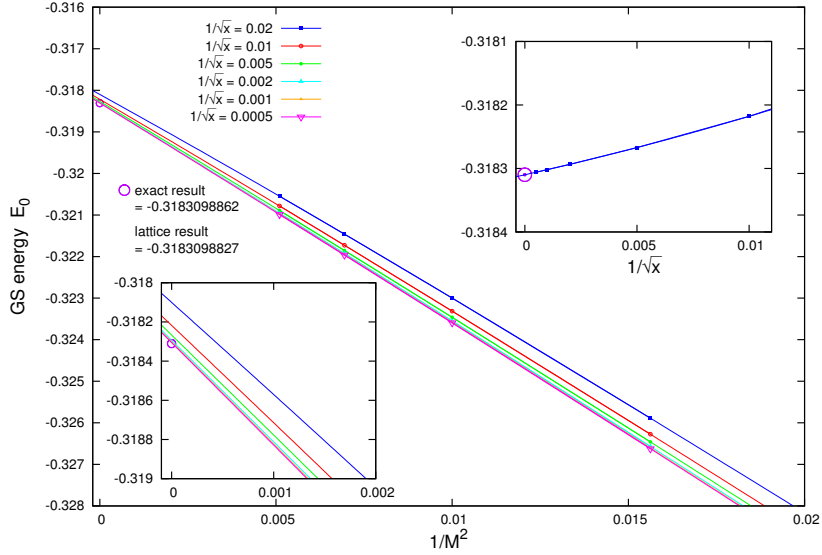}
\end{center}
\caption{\label{fig:E0} The main plot presents the dependence of our lattice data (at different
couplings $x$) for the GS energy $E_0$ on $1/M^2$, showing the extrapolation to the infinite volume
limit according to Eq.~\eqref{fit:vol}. The left inset is the close-up of the main plot in the
region of $M\rightarrow\infty$. The right inset shows the extrapolation of infinite volume
results at different couplings $1/\sqrt{x}$ according to Eq.~\eqref{fit:g}. Circles at the origin of
each plot show the exact infinite volume continuum result. Note that the point corresponding to
$x=4000000$ is plotted in the right inset, but it is not included in the continuum fit.}
\end{figure}

\begin{figure}[t!]
\begin{center}
\includegraphics
[width=0.6\textwidth,angle=270]
{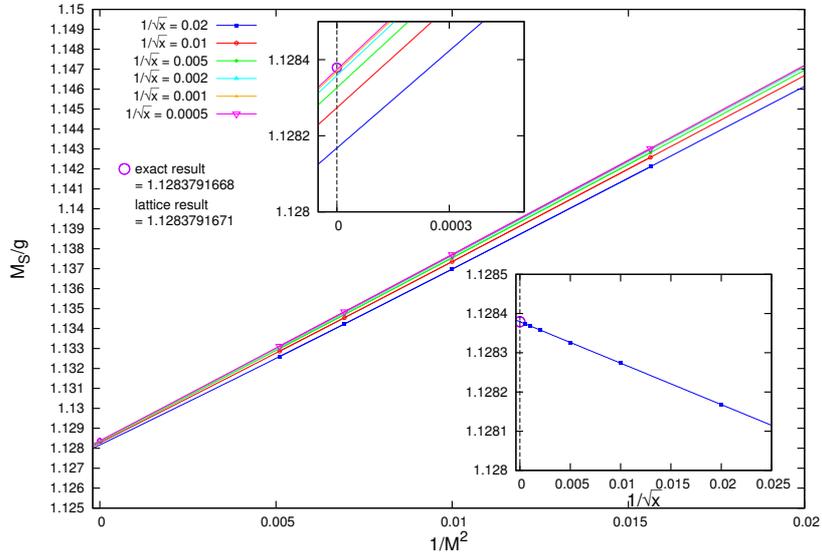}
\end{center}
\caption{\label{fig:scalar} As Fig.~\ref{fig:E0}, but for the scalar mass gap $M_S/g$.}
\end{figure}

\begin{figure}[t!]
\begin{center}
\includegraphics
[width=0.6\textwidth,angle=270]
{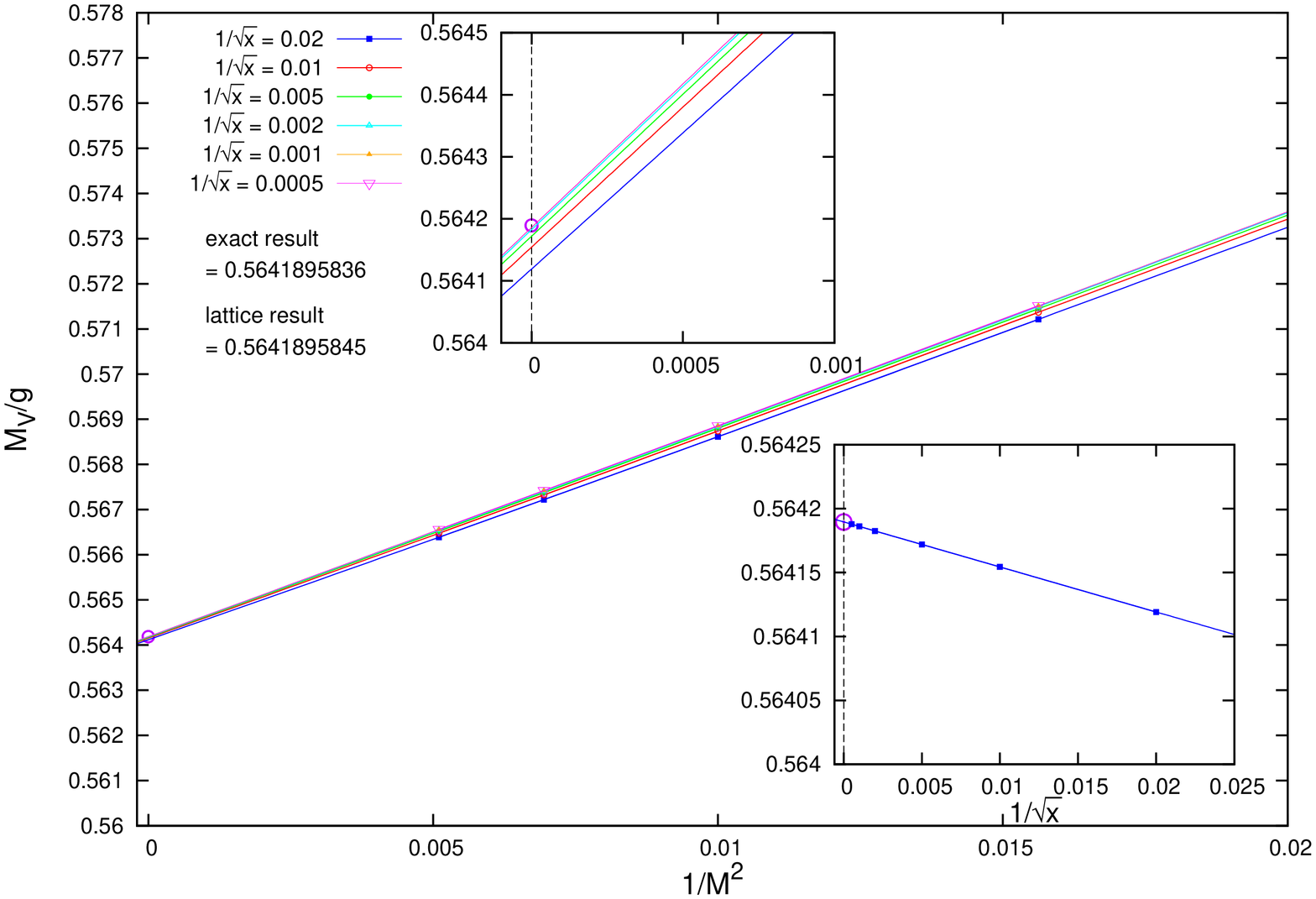}
\end{center}
\caption{\label{fig:vec} As Fig.~\ref{fig:E0}, but for the vector mass gap $M_V/g$.}
\end{figure}

We now discuss the extrapolation of our finite-lattice results to the infinite volume and continuum
limits.
As pointed out in Ref.~\cite{Hamer:1982mx}, it is important to take these limits in the correct order
-- first the infinite volume (bulk) limit and then the continuum limit.
It was suggested in Ref.~\cite{Crewther:1979ka} that the approach to the $M\rightarrow\infty$ limit
may be performed using the finite size scaling method of Fisher and Barber \cite{Fisher:1972zza}.
However, it was later realized that lattice data shows rather a power law behaviour close to the
continuum limit \cite{Hamer:1997dx}. In our data, we clearly find that this is indeed the case. The
origin of this
behaviour\footnote{Similar power law behaviour in small-volume QCD was observed in
Refs.~\cite{Fukugita:1992jj} and \cite{Fukugita:1992wq} and models were proposed therein to explain
it.} was attributed \cite{Hamer:1997dx} to the fact that the excitations have finite momentum
$\mathcal{O}(\pi/M)$ and their energies thus obtain a kinetic energy correction term
$\mathcal{O}((\pi/M)^2)$. Approaching the continuum limit $x\rightarrow\infty$, the power law
corrections become increasingly important.

\begin{table}[t!]
  \centering
  \begin{tabular}{cccc}
    \hline\hline
     & $E_0$ & $M_S/g$ & $M_V/g$\\
  \hline
  exact & -0.3183098862 & 1.1283791668 & 0.5641895836 \\
  \hline
  \multicolumn{4}{c}{QUADRATIC FITS IN $1/M^2$}\\
  $F_{00}$ & -0.3183103983 & 1.1283792781 & 0.5641898069\\
  $F_{01}$ & 0.009812 & -0.013058 & -0.004361 \\
  $F_{02}$ & 0.121528 & -- & -- \\
  error & $1.6\cdot 10^{-4}$\,\% & $9.8\cdot 10^{-6}$\,\% & $4.0\cdot 10^{-5}$\,\%\\
  \hline
  \multicolumn{4}{c}{CUBIC FITS IN $1/M^2$}\\
  $F_{00}$ & -0.3183098827 & 1.1283791671 & 0.5641895845\\
  $F_{01}$ & 0.00793666 & -0.010565 & -0.003522 \\
  $F_{02}$ & 0.126465 & -0.000144 & -- \\
  error & $1.1\cdot 10^{-6}$\,\% & $2.9\cdot 10^{-8}$\,\% & $1.8\cdot 10^{-7}$\,\%\\
    \hline\hline
  \end{tabular}
  \caption{Estimates of fitting coefficients in Eq.~\eqref{fit:g}. We quote values for 2 cases: the
infinite volume limit extrapolations of Eq.~\eqref{fit:vol} are performed either with quadratic or
cubic fits in $1/M^2$.
The values of $F_{00}|_{{\rm cont.},\,M=\infty}$ (denoted by $F_{00}$ for shortness) are our infinite
volume, continuum limit estimates and should be compared
with exact values. The error is $|(F_{00}-\textrm{exact})/\textrm{exact}|$. The quadratic corrections
in $ag$ ($F_{02}$) are statistically insignificant for the mass gaps (and hence not given), except for
the case of $M_S/g$ with
infinite volume taken including a cubic term in $1/M^2$.}
  \label{tab:fits}
\end{table}

We show our results for the extrapolation to the infinite volume limit in the main
plots of Fig.~\ref{fig:E0} (for the GS energy $E_0$), Fig.~\ref{fig:scalar} (for the scalar mass gap
$M_S/g$) and Fig.~\ref{fig:vec} (for the vector mass gap $M_V/g$). The left insets show the
neighbourhood of $1/M^2\rightarrow0$. In these plots, the circle at $1/M^2=0$ indicates the exact
infinite volume continuum result.
We fitted our data using the following polynomial ansatz (cubic in $1/M^2$) with fitting
parameters $F_0(ag)|_{M=\infty}$, $F_2$, $F_4$ and $F_6$\,:
\begin{equation}
\label{fit:vol}
 F(1/M^2) |_{ag} = F_0(ag)|_{M=\infty} + F_2\cdot\frac{1}{M^2} + F_4\cdot\frac{1}{M^4} +
F_6\cdot\frac{1}{M^6},
\end{equation} 
where $F(1/M^2) |_{ag}$ stands for either of the three considered quantities: $E_0$, $M_S/g$ or
$M_V/g$ at a fixed coupling $ag$ and for lattice size $M$.
Thus, the fitting parameter $F_0(ag)|_{M=\infty}$ is the infinite volume result at fixed coupling
(i.e. at non-zero lattice spacing). 
We also tried restricting the polynomial to a linear or quadratic one in $1/M^2$.
Obviously, the cubic fit has 4 parameters and we fit 4 data points -- hence the ``fits'' go exactly
through all points. However, our precision is such that the inclusion of the quadratic and then
the cubic term systematically improves our continuum limit value estimate.
Visually, the ``fits'' are straight lines in $1/M^2$ -- the quadratic and cubic corrections are
small, but still relevant from the point of extrapolation to the continuum limit.
Figs.~\ref{fig:E0}, \ref{fig:scalar} and \ref{fig:vec} show that the infinite volume limit of each
series of volumes approaches the continuum limit value as $x$ is increased. The
infinite volume values $F_0(ag)|_{M=\infty}$ for different couplings $ag\equiv1/\sqrt{x}$ are then
plotted against
$ag$ on the right insets of Figs.~\ref{fig:E0}, \ref{fig:scalar} and \ref{fig:vec}.
The fitting ansatz for the continuum limit $ag\rightarrow0$ extrapolation is:
\begin{equation}
\label{fit:g}
 F_0(ag)|_{M=\infty}=F_{00}|_{{\rm cont.},\,M=\infty} + F_{01}\cdot(ag) + F_{02}\cdot(ag)^2,
\end{equation} 
where the fitting parameters are $F_{00}|_{{\rm cont.},\,M=\infty}$, $F_{01}$, $F_{02}$.
In this way, the extracted values of $F_{00}|_{{\rm cont.},\,M=\infty}$ are the infinite
volume, continuum results that can be compared to analytical formulae.
The fitting curves for the mass gaps, shown in Figs.~\ref{fig:scalar} and \ref{fig:vec}, are to a
very good approximation straight lines in $ag$, whereas the curve for the GS energy,
Fig.~\ref{fig:E0}, shows some mild curvature.

The parameters of these fits are given in Tab.~\ref{tab:fits}. Indeed, the quadratic correction
$F_{02}\cdot(ag)^2$ is sizable only for $E_0$.
We observe systematic improvement of the compatibility of our final continuum limit results
$F_{00}|_{{\rm cont.},\,M=\infty}$ with the exact result when the term cubic in
$1/M^2$ is introduced in the fitting ansatz
of Eq.~\eqref{fit:vol}.
It constitutes a very small correction, but at this level of precision it improves the infinite volume
estimates $F_0(ag)|_{M=\infty}$ such that the final continuum estimates
$F_{00}|_{{\rm cont.},\,M=\infty}$ are closer to the exact result by
1-2 orders of magnitude\footnote{Introducing
the quadratic term in $1/M^2$ (in Eq.~\eqref{fit:vol}) is,
of course, even more important, improving the continuum limit result by 2-3 orders of magnitude.}.
The best precision is reached for the scalar mass gap, where our estimate is only $2.9\cdot
10^{-8}$\,\% off the exact value, i.e. the difference is on the tenth decimal place.
The results for the vector mass gap and GS energy are only slightly worse.

\section{Summary and discussion}
\label{sec:summary}
It is clear that the precision of the final results, in the simultaneous infinite volume limit
(taken first) and the continuum limit, can be arbitrarily increased.
By taking larger lattice sizes, one could introduce a term quartic
in $1/M^2$ (in Eq.~\eqref{fit:vol}), which would make it possible to observe sensitivity of
the continuum limit fit (Eq.~\eqref{fit:g}) to a term quadratic or cubic in $ag$ (for the mass gaps
or GS energy, respectively) and perform simulations even closer to the continuum limit. 

\begin{table}[t!]
  \centering
  \begin{tabular}{ccccc}
    \hline\hline
     & \multicolumn{2}{c}{$M_S/g$} & \multicolumn{2}{c}{$M_V/g$}\\
  & result & error & result & error\\
  \hline
  exact &  1.1283791668 & -- & 0.5641895836 & --\\
  \hline
  this work &  1.1283791671 & $2.9\cdot 10^{-8}$\,\% & 0.5641895845 &
$1.8\cdot 10^{-7}$\,\% \\
Ref.~\cite{Crewther:1979ka} &  1.120 & 0.7\,\% & 0.560 & 0.7\,\%\\
Ref.~\cite{Irving:1982yw}  & 1.128 & 0.03\,\% & 0.565 & 0.1\,\% \\
Ref.~\cite{Hamer:1997dx} (I)  & 1.25 & 11\,\% & 0.56 & 0.7\,\% \\
Ref.~\cite{Hamer:1997dx} (II)  & 1.14 & 1\,\% & 0.57 & 1\,\% \\
Ref.~\cite{Sriganesh:1999ws} (I)  & 1.11 & 1.6\,\% & 0.563 & 0.2\,\%\\
Ref.~\cite{Sriganesh:1999ws} (II)  & 1.1284 & 0.002\,\% & 0.56417 & 0.003\,\%\\
Ref.~\cite{Byrnes:2002nv}  & -- & -- & 0.56419 & $7\cdot10^{-5}$\,\% \\
    \hline\hline
  \end{tabular}
  \caption{Comparison of results for the scalar and vector mass gaps presented in this work and in
selected literature.}
  \label{tab:compare}
\end{table}

Obviously, since exact results are known, this is rather pointless\footnote{Note also that at our
current level of precision for the quantities of interest, one is approaching the problem
of machine precision, since typical algorithms for finding eigenvalues can give their magnitudes
accurate to
$\mathcal{O}(10^{-10})$ with double precision numbers (several iterations of an algorithm with many
operations per iteration, accumulating $\mathcal{O}(10^{-15})$ errors per operation). Hence,
increasing precision beyond what we present in this paper would need an increased machine precision.}.
We believe, though, that the results presented in this paper are very interesting.
Exact results are approached with remarkable rapidity using only moderate lattice sizes -- 8 to 14
lattice sites.
The success of the method should therefore be attributed to working very close to the continuum
limit, at inverse coupling $x\equiv\beta$ a few orders of magnitude larger than in typical lattice
simulations, using e.g. Monte Carlo methods.
In earlier applications of Hamiltonian techniques, it was also not realized that the
continuum limit extrapolation can be performed using very high values of $x$.
Using rather small values of $x$, exact results could be approached with a precision of the
order of 0.1-1\,\% (see Tab.~\ref{tab:compare}). The impressive improvement of this precision with the
DMRG method \cite{Byrnes:2002nv} by 2-3 orders of magnitude\footnote{In Tab.~\ref{tab:compare},
the error of the final result for the vector mass gap is $7\cdot10^{-5}\,\%$. However, taking into
account the estimated error, i.e. $M_V/g=0.56419(4)$, the actual precision claimed by the Authors of
Ref.~\cite{Byrnes:2002nv} is of $\mathcal{O}(10^{-3})\,\%$.} seemed to set impassable limits for
lattice Hamiltonian methods.
However, as we have shown in this paper, this precision can be easily surpassed by another 3-4
orders of magnitude, applying only very simple ideas.

There is no straightforward way to extend the method presented in this paper to the
massive case. The reason for this is the following.
The mass gap $M_{\mu}/g$ ($\mu=V,\,S$) at an arbitrary quark mass $m/g$ can be expressed as:
\begin{equation}
\label{eq:massgap}
 \frac{M_{\mu}}{g}\left(\frac{m}{g}\right) = \frac{M_{\mu}^{(0)}}{g} + \delta\,\frac{
M_{\mu}}{g}\left(\frac{m}{g}\right),
\end{equation} 
where the first term $M_{\mu}^{(0)}/g$ is mass independent and given by the leading order term of mass
perturbation theory of mass gap expansion \cite{Frohlich:1976mt,Adam:1995us,Vary:1996uc}, i.e.
$1/\sqrt{\pi}$ or $2/\sqrt{\pi}$ for the vector and scalar case, respectively.
The second term $\delta M_{\mu}/g\,(m/g)$, the mass gap shift from massive quarks, depends on $m/g$
and has been calculated in mass perturbation theory up to second order in $m/g$.
Our data indicate that $M_{\mu}^{(0)}/g$ can be calculated accurately even if the physical system size
is very small with respect to the size of the meson under consideration (which is the case in our
analysis very close to the continuum limit, i.e. at very small lattice spacing, and with our rather
small lattice sizes).
However, the shift $\delta M_{\mu}/g\,(m/g)$, incorporating effects of quark masses, requires that the
physical system size is large with respect to meson size.
Our small physical system sizes are therefore large enough to extract $M_{\mu}^{(0)}/g$ and finite
size effects for this quantity are accurately described by Eq.~\eqref{fit:vol}.
They are not enough to describe the effects of massive quarks, though.
This indicates that the nature of finite size effects is different for the two terms in
Eq.~\eqref{eq:massgap}.
Indeed, a naive application of the method described in this paper for $m/g\neq0$ yields values for
the mass gaps very close to the ones in the massless case, i.e. $M_{\mu}^{(0)}/g$ is reproduced
correctly, but $\delta M_{\mu}/g\,(m/g)$ is not.

A possible solution is then to keep away from the continuum limit (work at much smaller $x$),
such that the lattice spacing is coarse enough and the physical extent of the lattice is large with
respect to the meson size.
This allows to truncate SCE at some small order $N\approx M$ (with no or very few flux loops),
which is then enough to saturate the ground state energy and mass gaps. This, however,
makes it necessary to perform a long extrapolation to the continuum limit and the amazing precision of
the final result is lost.
Our analysis of the massive case will be presented elsewhere.

\vspace*{0.5cm}
\noindent\textbf{Acknowledgments.} We acknowledge useful discussions with Mari Carmen Ba\~nuls,
Karl Jansen, Piotr Korcyl and Piotr Tomczak. K.C. was supported by Foundation for Polish Science
fellowship ``Kolumb''.


\begin{thebibliography}{99}

\bibitem{Schwinger:1962tp}
  J.~S.~Schwinger,
  Phys.\ Rev.\  {\bf 128} (1962) 2425.

\bibitem{Gutsfeld:1999pu}
  C.~Gutsfeld, H.~A.~Kastrup and K.~Stergios,
  Nucl.\ Phys.\ B {\bf 560} (1999) 431
  [hep-lat/9904015].

\bibitem{Gattringer:1999gt}
  C.~Gattringer, I.~Hip and C.~B.~Lang,
  Phys.\ Lett.\ B {\bf 466} (1999) 287
  [hep-lat/9909025].

\bibitem{Giusti:2001cn}
  L.~Giusti, C.~Hoelbling and C.~Rebbi,
  Phys.\ Rev.\ D {\bf 64} (2001) 054501
  [hep-lat/0101015].

\bibitem{Christian:2005yp}
  N.~Christian, K.~Jansen, K.~Nagai and B.~Pollakowski,
  Nucl.\ Phys.\ B {\bf 739} (2006) 60
  [hep-lat/0510047].

\bibitem{Bietenholz:2011ey}
  W.~Bietenholz, I.~Hip, S.~Shcheredin and J.~Volkholz,
  Eur.\ Phys.\ J.\ C {\bf 72} (2012) 1938
  [arXiv:1109.2649 [hep-lat]].

\bibitem{Banks:1975gq}
  T.~Banks, L.~Susskind and J.~B.~Kogut,
  Phys.\ Rev.\ D {\bf 13} (1976) 1043.

\bibitem{Susskind:1975hj}
  L.~Susskind and J.~B.~Kogut,
  Phys.\ Rept.\  {\bf 23} (1976) 348.

\bibitem{Carroll:1975gb}
  A.~Carroll, J.~B.~Kogut, D.~K.~Sinclair and L.~Susskind,
  Phys.\ Rev.\ D {\bf 13} (1976) 2270
   [Erratum-ibid.\ D {\bf 14} (1976) 1729].

\bibitem{Kenway:1977dk}
  R.~D.~Kenway and C.~J.~Hamer,
  Nucl.\ Phys.\ B {\bf 139} (1978) 85.


\bibitem{Crewther:1979ka}
  D.~P.~Crewther and C.~J.~Hamer,
  Nucl.\ Phys.\ B {\bf 170} (1980) 353.

\bibitem{Jones:1979av}
  D.~R.~T.~Jones, R.~D.~Kenway, J.~B.~Kogut and D.~K.~Sinclair,
  Nucl.\ Phys.\ B {\bf 158} (1979) 102.


\bibitem{Hamer:1982mx}
  C.~J.~Hamer, J.~B.~Kogut, D.~P.~Crewther and M.~M.~Mazzolini,
  Nucl.\ Phys.\ B {\bf 208} (1982) 413.


\bibitem{Irving:1982yw}
  A.~C.~Irving and A.~Thomas,
  Nucl.\ Phys.\ B {\bf 215} (1983) 23.

\bibitem{Schiller:1983sj}
  A.~Schiller and J.~Ranft,
  Nucl.\ Phys.\ B {\bf 225} (1983) 204.

\bibitem{Fang:1992bi}
  X.~- Y.~Fang, X.~- Q.~Luo, G.~- C.~Xu and Q.~- Z.~Chen,
  Z.\ Phys.\ C {\bf 54} (1992) 587.


\bibitem{Aroca:1997hp}
  J.~M.~Aroca, H.~Fort and G.~Alvarez,
  Europhys.\ Lett.\  {\bf 45} (1999) 565
  [hep-lat/9711049].


\bibitem{Hamer:1997dx}
  C.~J.~Hamer, W.~-h.~Zheng and J.~Oitmaa,
  Phys.\ Rev.\ D {\bf 56} (1997) 55
  [hep-lat/9701015].

\bibitem{Kroger:1998se}
  H.~Kroger and N.~Scheu,
  Phys.\ Lett.\ B {\bf 429} (1998) 58
  [hep-lat/9804024].

\bibitem{Sriganesh:1999ws}
  P.~Sriganesh, R.~Bursill and C.~J.~Hamer,
  Phys.\ Rev.\ D {\bf 62} (2000) 034508
  [hep-lat/9911021].

\bibitem{Fang:2001gq}
  X.~-Y.~Fang, D.~Schutte, V.~Wethkamp and A.~Wichmann,
  Phys.\ Rev.\ D {\bf 64} (2001) 014501.


\bibitem{Byrnes:2002nv}
  T.~Byrnes, P.~Sriganesh, R.~J.~Bursill and C.~J.~Hamer,
  Phys.\ Rev.\ D {\bf 66} (2002) 013002
  [hep-lat/0202014].



\bibitem{Kogut:1976}
  J.~B.~Kogut and L.~Susskind,
  Phys.\ Rev.\ D {\bf 11} (1975) 395.


\bibitem{jw}
  P.~Jordan, E.~Wigner,
  Z.\ Phys.\ {\bf 47} (1928) 631.
  
\bibitem{Fisher:1972zza}
  M.~E.~Fisher and M.~N.~Barber,
  Phys.\ Rev.\ Lett.\  {\bf 28} (1972) 1516.


\bibitem{Fukugita:1992jj}
  M.~Fukugita, H.~Mino, M.~Okawa, G.~Parisi and A.~Ukawa,
  Phys.\ Lett.\ B {\bf 294} (1992) 380.

\bibitem{Fukugita:1992wq}
  M.~Fukugita, H.~Mino, M.~Okawa, G.~Parisi and A.~Ukawa,
  Nucl.\ Phys.\ Proc.\ Suppl.\  {\bf 30} (1993) 365.


\bibitem{Frohlich:1976mt}
  J.~Fr\"ohlich and E.~Seiler,
  Helv.\ Phys.\ Acta {\bf 49} (1976) 889.

\bibitem{Adam:1995us}
  C.~Adam,
  Phys.\ Lett.\ B {\bf 382} (1996) 383
  [hep-ph/9507331].

\bibitem{Vary:1996uc}
  J.~P.~Vary, T.~J.~Fields and H.~-J.~Pirner,
  Phys.\ Rev.\ D {\bf 53} (1996) 7231.



\end{thebibliography}
\end{document}